# Use Bagging Algorithm to Improve Prediction Accuracy for Evaluation of Worker Performances at a Production Company

Hamza R Saad*

*Department of Industrial Engineering and System Science, Binghamton University, USA*

## Abstract

Many workers at the production department of Libyan Textile Company work with different performances. Plan of company management is paying the money according to the specific performance and quality requirements for each worker. Thus, it is important to predict the accurate evaluation of workers to extract the knowledge for management, how much money it will pay as salary and incentive. For example, if the evaluation is average, then management of the company will pay part of the salary. If the evaluation is good, then it will pay full salary, moreover, if the evaluation is excellent, then it will pay salary plus incentive percentage. Twelve variables with 121 instances for each variable collected to predict the evaluation of the process for each worker. Before starting classification, feature selection used to predict the influential variables which impact the evaluation process. Then, four algorithms of decision trees used to predict the output and extract the influential relationship between inputs and output. To make sure get the highest accuracy, ensemble algorithm (Bagging) used to deploy four algorithms of decision trees and predict the highest prediction result 99.16%. Standard errors for four algorithms were very small; this means that there is a strong relationship between inputs (7 variables) and output (Evaluation). The curve of (Receiver operating characteristics) for algorithms gave a high- level specificity and sensitivity, and Gain charts were very close to together. According to the results, management of the company should take a logic decision about the evaluation of production process and extract the important variables that impact the evaluation.

## Introduction

Data mining has a vital role in predicting the roadmap of production. Many companies entered data mining as a tool to solve complex problems and extract the knowledge from vast and vague data. The manufacturing process is very complicated to understand by traditional techniques, so statistical analysis and quality tools are unable to handle all daily data. Thus, data mining is a proper technique uses to extract a basic knowledge to build the relationship between variables and take right decision to improve evaluation process [1]. This case study related to the production process for predicting the performance of each hard worker based on practical input variables.

Few studies focused on worker evaluation, especially in production and manufacturing. Searchers used data mining methodology to extract essential patterns from the institutional database [2]. Different data algorithms such as association rules and K-means applied to predict evaluation performance. Current performance evaluation needs to support the recommendations for merit salary adjustments and in grade or grade change salary increases [3]. Also, that helps the supervisors to find the right performance which meets right salary, and those employees who have excellent performance need particular attention to optimize falling ratio by taking action at a specific time. A decision tree is applied by focusing on particular variables which impact the final process. Prediction of whole evaluation related to selected data [4]. Some variables give weak impact, such as Age attribute did not present significant effect while Marital Status and Gender presented important prediction for performance evaluation.

Many algorithms used in the industrial field but only decision trees gave a sophisticated result whether for regression or classification, because that we used four algorithms of decision trees to predict production data, so, ensemble learning (bagging algorithm) used to deploy all algorithms to get one result and improved prediction.

Decision trees had good performance to handle linear or nonlinear data. However, one algorithm has the limited performance for prediction. Many algorithms used to classify production data, but most of the results were low accuracy and high standard error. Therefore, the decision tree has confirmed in the study to solve data because of many advantages:

1. A decision tree is performing feature selection or variable screening.
2. A decision tree is requiring little efforts by users for preparing data.
3. Nonlinear relationships between variables do not affect the tree performance.
4. It is easy to explain and interpret.

## Data Collection

Data collected from Libyan Textile Company for 12 attributes and 121 instances. Dependent or output is an evaluation; it selected as the measure of the performance of each worker.

### Variables or inputs are as following

1. Operator: A worker who is responsible for the production and manufacturing processes.

**\*Corresponding author:** Hamza R Saad, Department of Industrial Engineering and System Science, Binghamton University, USA, Tel: 6077772171; E-mail: hsaad1@binghamton.edu

2. Badge No: A reference number which gave to each worker.
3. Job title: A specific job should accomplish in production.
4. Base production: The minimum quantity of production should achieve to get a full salary.
5. Production achieved: Real production quantity.
6. Incentive wages: Money gives if worker exceeded base production.
7. Production rate: The ratio between real production and base production.
8. Labor efficiency: Percentage performance gave to each worker by responsible supervisor (the evaluation of production supervisor did not confirm by top management yet).
9. Machine: There are two types of the machine at the company, first old machine (high quality and slow productivity), and second, new machine (fast productivity and medium quality).
10. Product: Company produces five types of carpets, but the most important carpet is a woven carpet.
11. Elapsed time: A specific time wanted to produce a specific quantity.
12. The unit: Company has seven units, each unit designed for achieving a part of the job.

**The dependent variable is an evaluation**

1. If worker achieved less than base production, then management will pay only part of the salary.
2. If worker achieved base production, then management will pay full salary.
3. If worker achieved more than base production, then management will pay full salary plus incentives.

Figure 1 presents the input variables which applied for predicting the evaluation.

## Methodology

Many algorithms of data mining gave a low accuracy, and many records or instances were misclassified. However, decision trees gave the highest accuracy among applied algorithms. Before starting classification, four variables removed using feature selection as shown in Table 1. These attributes reduce the accuracy of prediction and give low impact on the output. Four decision trees applied for classification, then bagging or voting algorithm used to deploy all algorithms of decision trees to give one prediction result. Figure 2 shows a methodology of the study.

### Voting or bagging algorithm (Ensemble learning)

Ensemble Method started around ten years ago as a separate field of machine learning, and it enhanced by the idea of needing to increase the power for multiple algorithms and not only trust one algorithm built on the small training set. Experimental developments and important theoretical conducted for ten years led to several techniques, especially boosting and bagging, became used to solve many complicated problems. However, ensemble method also appeared to be applicable for now and future problems of online applications and distributed data mining [5].

Bootstrap Aggregating (Bagging) generates many bootstrap training sets from the original training set (using sampling with replacements) and using each of them for generating a classifier for including in the ensemble. The models for sampling with replacement and bagging presented in Figure 3. For these models. (T), it is the essential training set for (N) example. (M), it is the number of the base model to be learned. (Lb), it is the base model learning. The hi's are the base models. Random_integer of (a,b) is the function that returns every integer from a to b with the equal probabilities. Moreover, (I(A)) is the indicator function that returns 1 if A was true and 0 otherwise [6].

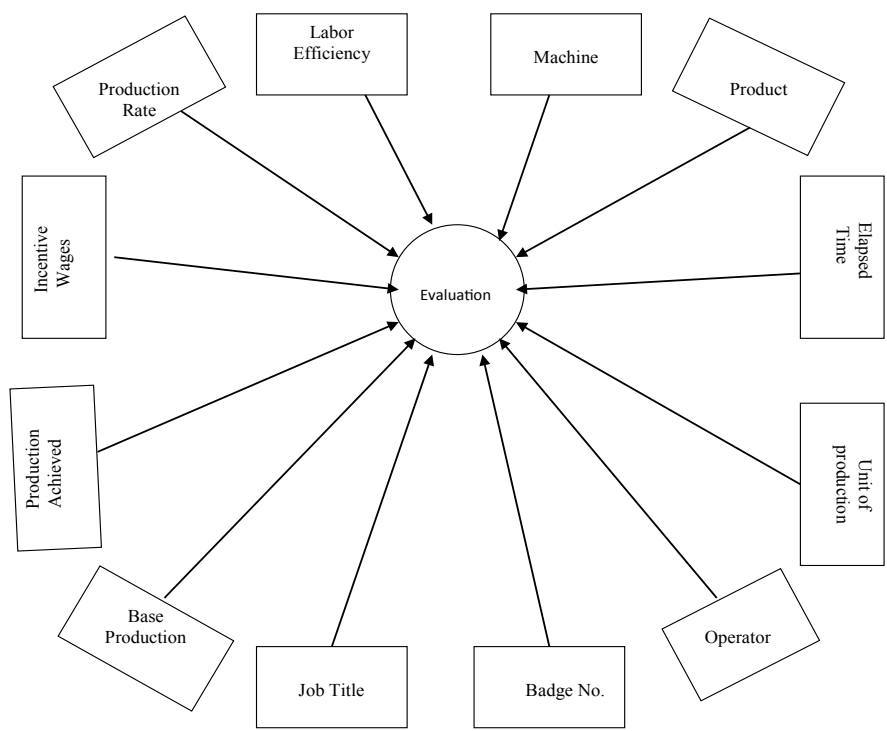

**Figure 1:** Input variables classified according to the evaluation process.

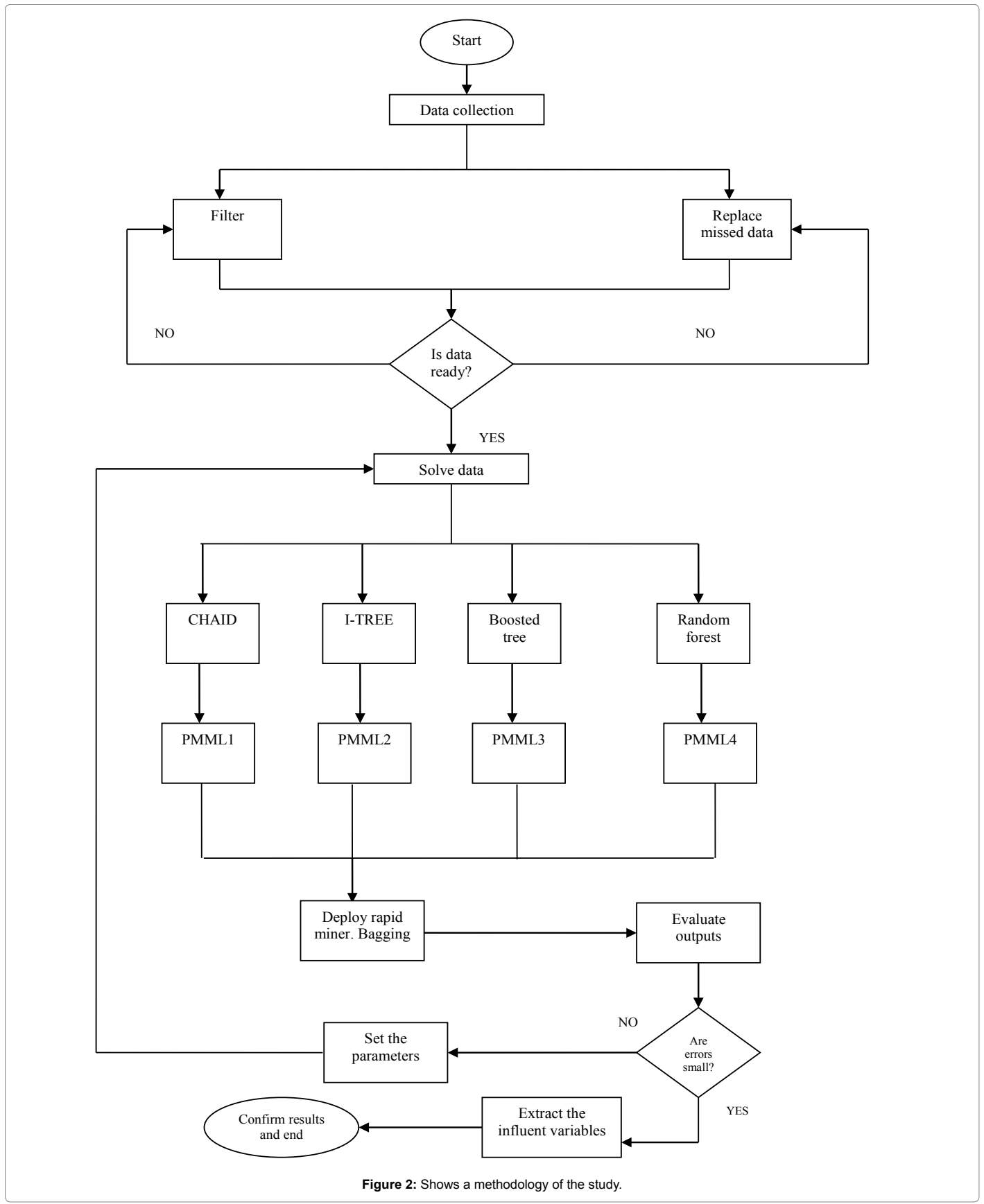

**Figure 2:** Shows a methodology of the study.

**The study used four types of decision trees**

1. Random forests trees.
2. Boosted trees.
3. Interactive tree (CART and CHAID). (Where the CART is Classification and Regression Tree, and (CHAID is Chi-squared Automatic Interaction Detector).
4. CHAID tree. (Where CHAID is Chi-squared Automatic Interaction Detector).

**Feature selection**

Select features use to increase the accuracy by ignoring the low-efficiency variables which lead to reduce the accuracy of data classification. Many algorithms applied in this study, but the accuracy did not exceed 65% and 71%, so this accuracy does not help enough to take the right decision. We believed that some attributes did not work well for prediction and should remove from the dataset. According to the feature selection result, four attributes removed (unit, product, elapsed time, and machine). Feature selection shows in Table 1. Variables removed based on p-value and chi-square values.

Until this point, management cannot predict the right decision for who deserves part of the salary, full salary, or full salary plus incentives because there is no clear relationship or interaction between variables. Algorithms of decision tree will predict the relationship and interactions based on evaluation levels [7-10].

**Results and Discussion**

Four decision trees gave high accuracy, and their standard error is very small, which help to accept results of each algorithm. The strategy of the decision tree depends on split data until getting a pure subset. From results, each algorithm gets a high probability of pure subset. Table 2 shows a small error rate got from each decision tree.

Data classified according to three levels, Average, Good, and Excellent. Most of the data classified correctly and very small data were miss-classified. Data for 121 workers, each worker needs to get a correct evaluation. Using bagging or voting algorithm, we got a sophisticated classification as follows:

Evaluation for Average, the total average was 27 and voted predicted average 27. Evaluation for Good, total good was 62 and voted predicted good 62.

Evaluation for Excellent, total excellent was 30 and voted predicted excellent 29; only one instance miss-classified.

Bagging algorithm as shown in the table below reached the highest accuracy 99.16% by deploying all decision trees in one model to give one predicted result.

**Summary frequency table**

In this table, we split the accuracy according to each level (Average, Good, Excellent) to show which level has the highest prediction according to the frequency.

**Gain harts**

The gain chart is the ratio of accurate predictions to the total number of that category response at the different percentiles. It shows a model performance compared to the baseline; it gives the indications of algorithm performance. So, gain chart shows the percentage of observations correctly classified for the given category as Average, Good, and Excellent (Figure 4).

In the Gain chart, we need to maximize the space between curves and baseline. For Average and Excellent classification, CHAID model gives the highest space between baseline and curve, whereas, for Good classification, EXHAUSTIVE CHAID model gives the highest space between baseline and curve. Therefore, the best performance came from CHAID and EXHAUSTIVE CHAID models (Table 3).

**Receiver operating characteristics (ROC curve)**

All decision trees in study give an excellent test because of all values are under the curve between 0.997585 and 1. CHAID model gives a high-level specificity and sensitivity for all classification levels. Areas under the curve for Average, Good, and Excellent classifications are equal to 1. Whereas, Exhaustive CHAID model only gives the highest specificity and sensitivity at Good classification equals to 1. After that, the high value of sensitivity and specificity came from Random forest model, values under the curve are 0.997585, 0.998585, 0.997753 respectively. Then, boosted tree model gives area under the curve less than rest of models, but it also has a high sensitivity and specificity (Figure 5).

|  | Chi-Square | P-value | Variable Number |
|---|---|---|---|
| Operator | 242.0000 | 0.000000 | 1 |
| Job Title | 30.1333 | 0.000000 | 3 |
| Badge No. | 55.6470 | 0.000000 | 2 |
| Production rate | 133.9668 | 0.000000 | 7 |
| Labor efficiency | 187.5764 | 0.000000 | 8 |
| Base Production | 51.2168 | 0.000000 | 4 |
| Incentive Wages | 83.2093 | 0.000000 | 6 |
| Production Achieved | 27.1449 | 0.000140 | 5 |
| Unit | 11.7627 | 0.301250 | 12 |
| Product | 6.4121 | 0.601180 | 10 |
| Elapsed time | 6.3258 | 0.610790 | 11 |
| Machine | 0.7942 | 0.672260 | 9 |

**Table 1:** Feature selection.

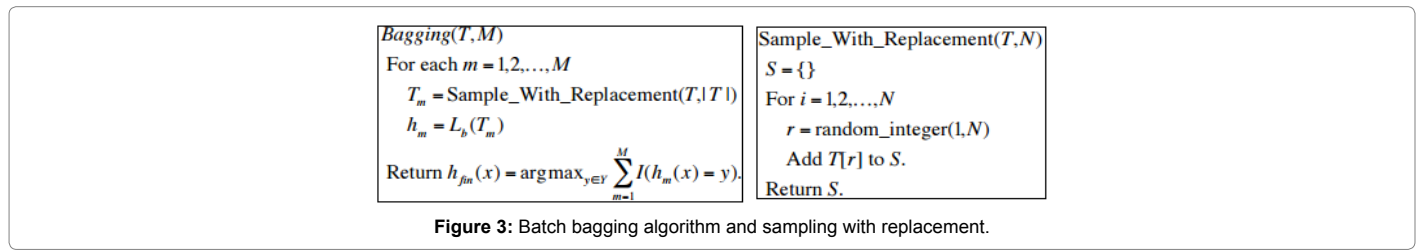

**Figure 3:** Batch bagging algorithm and sampling with replacement.

|  | Boost Tree Model | CHAID Model | Exhaustive CHAID Model | Random Forest Model |
|---|---|---|---|---|
| Error rate | 0.025210 | 0.00 | 0.025210 | 0.042017 |

**Table 2:** Error rate for each algorithm.

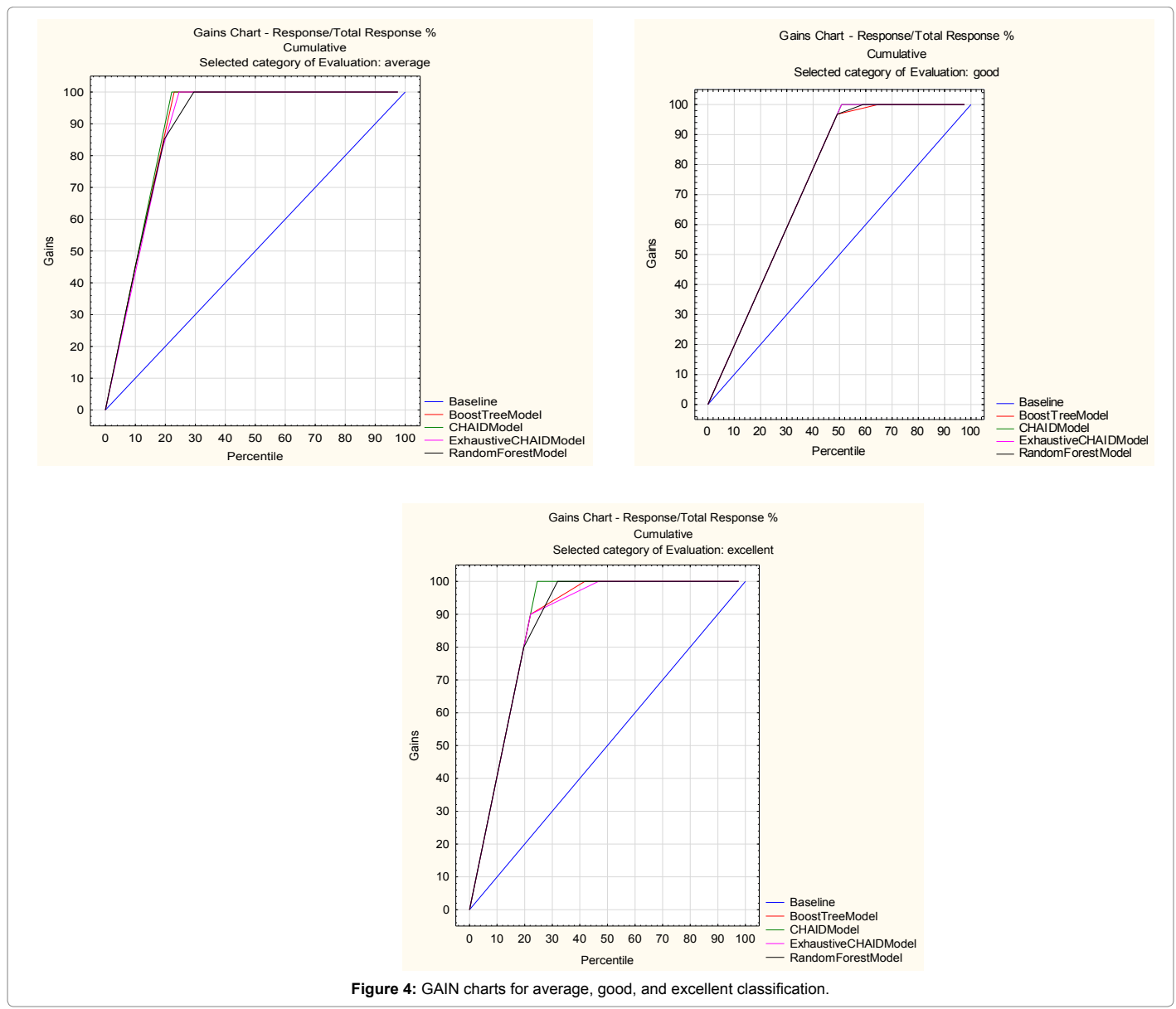

**Figure 4:** GAIN charts for average, good, and excellent classification.

|  | Evaluation | Voted predicted Average | Voted predicted Good | Voted predicted Excellent | Row Total |
|---|---|---|---|---|---|
| Count | Average | 27 | 0 | 0 | 27 |
| Column percent |  | 96.43% | 0.00% | 0.00% |  |
| Row percent |  | 100.00% | 0.00% | 0.00% |  |
| Total percent |  | 22.69% | 0.00% | 0.00% | 22.69% |
| Count | Good | 0 | 62 | 0 | 62 |
| Column percent |  | 0.00% | 100.00% | 0.00% |  |
| Row percent |  | 0.00% | 100.00% | 0.00% |  |
| Total percent |  | 0.00% | 52.10% | 0.00% | 52.10% |
| Count | Excellent | 1 | 0 | 29 | 30 |
| Column percent |  | 3.57% | 0.00% | 100.00% |  |
| Row percent |  | 3.33% | 0.00% | 96.67% |  |
| Total percent |  | 0.84% | 0.00% | 24.37% | 25.21% |
| Count | All Grps | 28 | 62 | 29 | 119 |
| Percent |  | 23.53% | 52.10% | 24.37% |  |

**Table 3:** Summary of the frequency.

Table 4 shows the area under the curve for different classification levels using different decision tree models

All models of decision trees did an excellent job by getting high accuracy, and this accuracy helps bagging algorithm to give a high accuracy as well by building new accuracy according to previous model's accuracies. The accuracy 99.16% is exciting to take a right decision by top management.

Table 5 shows the accuracy of each algorithm of decision trees and Bagging algorithm as well.

Values of accuracy are between 94.21% and 98.35%, and Bagging algorithm developed the accuracy to 99.16%. Therefore, ensemble learning is a good strategy used to improve the accuracy, but to improve accuracy we should focus on these points:

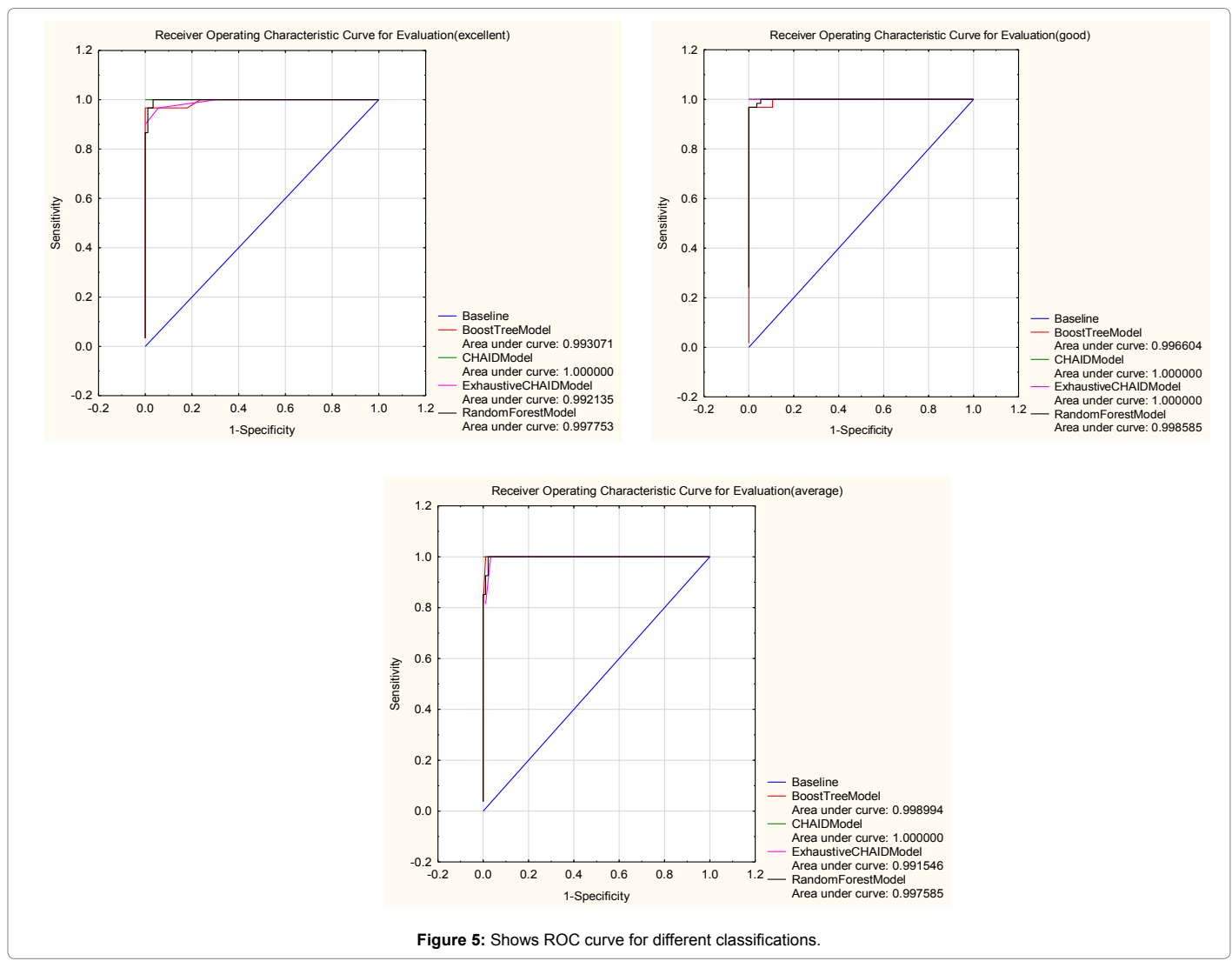

**Figure 5:** Shows ROC curve for different classifications.

| The area under the curve | | | |
|---|---|---|---|
| ROC curve for algorithms | **Average** | **Good** | **Excellent** |
| Boost tree model | 0.996994 | 0.996604 | 0.993071 |
| CHAID model | 1 | 1 | 1 |
| Exhaustive CHAID model | 0.991545 | 1 | 0.992135 |
| Random forest model | 0.997585 | 0.998585 | 0.997753 |

**Table 4:** ROC results of the area under the curve.

| Algorithm | Overall Accuracy |
|---|---|
| Random forest | 94.21% |
| Boosted tree | 95.87% |
| Interactive tree | 95.87% |
| CHAID tree | 98.35% |
| Voted or Bagging | 99.16% |

**Table 5:** Accuracy for each algorithm.

1. Select a robust algorithm able to handle all dataset.
2. Remove weak variables which have a weak relationship with the output.
3. Preprocess data using a suitable filter if data has any noisy or weak correlation.
4. Use ensemble learning based on suitable algorithms.

## Conclusion and Future Work

Management of company can build robust prediction for each worker and pay right money for evaluation by focusing on critical factors. In the study, seven variables used to predict final evaluation; Operator, Job Title, Badge No, Production rate, Labor efficiency, Base Production, Incentive Wages, and Production Achieved. Whereas four variables were ignored using screening feature, variables are; Unit, Product, Elapsed time, and Machine.

Four algorithms gave a high accuracy result and very small errors rate. In the study, a strong relationship between variables able to give a sophisticated evaluation. Bagging algorithm gives a very high accuracy, but this accuracy cannot be achieved without using suitable algorithms like decision trees. CHAID model gave a very high-level specificity and sensitivity, so it gives the highest space between baseline and model curve. According to the results, management can take a right decision by concentrating on the influential variables that gave a very small error rate. For instance, the evaluation is Excellent based on Production rate, Job title, Labor efficiency, and Base production variables. Evaluation is Average and good based on Production rate, Labor efficiency, and Badge No. In future work; we need to collect different type of non-linear data and apply more data mining algorithm beside machine learning to get more relationship between variables from the complicated production process.